\documentstyle[preprint,prb,aps]{revtex}
\begin{document}
\draft
\title{Many skyrmion wave functions and skyrmion statistics
in quantum Hall ferromagnets}

\author{Kun Yang and S. L. Sondhi}

\address{Department of Physics, Princeton University,
Princeton, NJ 08544}
\date{\today}

\maketitle

\begin{abstract}

We determine the charge and statistical angle of skyrmions in quantum Hall 
ferromagnets by performing Berry phase calculations based on the 
microscopic variational wave functions for many-skyrmion states. We find,
in contradiction to a recent claim by Dziarmaga, that
both the charge and the statistical
angle of a skyrmion are {\em independent} of  
its spin (size), and are identical to those of Laughlin quasiparticles
at the same filling factor. 
We discuss some subtleties in the use of these variational wave functions.

\end{abstract}


Recently there has been considerable interest in quantum Hall 
systems that spontaneously develop spin 
ferromagnetism,\cite{sondhi} as well as in double layer quantum
Hall systems with spontaneous interlayer phase coherence,
where the layer degrees of freedom play a role similar to that of
spins of the electrons.\cite{yang,moon}
Of special interest is the realization that the low energy
charged excitations in these systems are highly collective topological
solitons called skyrmions.\cite{leekane,sondhi,fertig,theory,macd} 
It is now well understood that
the charge carried by a skyrmion is equal to its topological charge times the
Landau level filling factor of the system, while its energy and spin 
(number of electrons whose spins are flipped in the core region of the
skyrmion) depend
sensitively on the strength of Zeeman field.\cite{sondhi,fertig} These
theoretical predictions have been confirmed experimentally for
the case of filling factor $\nu=1$.\cite{barrett,eisenstein,aifer}

In reformulating the low energy physics near ferromagnetic fillings
in terms of skyrmions, the issue of what statistics to assign them
becomes important. 
While several previous authors have addressed this starting from the
bosonic Chern-Simons approach to the quantum Hall 
effect\cite{zhang,leekane}, consensus has not been reached. 
Sondhi {\em et al.}\cite{sondhi} and Nayak and Wilczek \cite{nayak}
have argued that there is a
Hopf term in the effective $\sigma$ model that describes the low
energy spin dynamics of the system, which enforces, for the skyrmions, 
statistics identical to those of the polarized, Laughlin 
quasiparticles\cite{laughlin} at the same filling factor.
The same conclusion was reached by Moon {\em et al.} based on 
an analysis of
the $CP^1$ model and duality transformations.\cite{moon}
Recently however, Dziarmaga\cite{dz} has claimed that the statistical 
angle of the skyrmions depends on their spin (or core size). Were
this to be the case, the consequences would be striking: both even
and odd denominator states would arise from the standard hierarchy 
construction carried out with skyrmions, and their realization in a 
given sample would depend sensitively on the value of the Zeeman
energy.

In this paper we show that this does not happen and that both the
charge and the statistical angle of a skyrmion depend {\em only} on 
its topological properties, and are identical to those of the Laughlin 
quasiparticles at the same filling factor. We show this by
performing direct Berry phase calculations with generalizations of
the {\em microscopic} variational wave functions for skyrmions proposed 
by Moon {\em et al.},\cite{moon} in close
analogy to the approach taken by Arovas {\em et al.}\cite{arovas} in 
computing the charge and statistics of Laughlin quasiparticles in the
spin polarized quantum Hall effect. 
Before turning to the details of the calculations we discuss the
wave functions appropriate to a system with many skyrmions.

\noindent
{\bf Many Skyrmion Wave Functions:}
In the symmetric gauge, the ground state of a quantum Hall system at filling
factor $\nu=1/m$ (where $m$ is an odd integer)
is well approximated by the wave function:
\begin{equation}
\psi_g(\chi_1, z_1, \cdots, \chi_N, z_N)
=\prod_{i=1}^N \left(\begin{array}{c} 1 \\ 0 \\ \end{array}
\right)_i \psi_m(z_1,\cdots,z_N),
\label{ground}
\end{equation}
where
\begin{equation}
\psi_m(z_1,\cdots,z_N)=
\prod_{j<k}^N(z_j-z_k)^m\exp(-\sum_{l}^N|z_l|^2/4\ell^2)
\label{laugh}
\end{equation}
is the Laughlin wave function that involves only the spatial coordinates of
the electrons.
Here $\chi_j$ is the spin wave function of the $j$th electron,
$z_j=x_j+iy_j$ is the complex coordinate of the $j$th electron,
$\ell$ is the magnetic length, and
$N$ is the total number of electrons in the system. It is the exact 
ground state wave
function for $m=1$ within the lowest Landau level (LLL) approximation.
We have assumed that the Zeeman field 
has polarized the spins of all electrons into the up direction. In the 
absence of Zeeman coupling, the electron spins are fully polarized
spontaneously
but the polarization vector may point in any direction. The system has a
macroscopic ground state degeneracy in this case. 

Variational wavefunctions for states with arbitrary distributions of 
skyrmions can be written down
as appropriate transcriptions of the known time independent classical 
solutions of the 2+1 dimensional scale invariant $O(3)$ non-linear $\sigma$ 
model.\cite{raja}  
Recall that the latter can be written in the unnormalized, spinorial form 
${f \choose g}$, where $f$ and $g$ are analytic functions of the 
complex two dimensional co-ordinate $z$. The corresponding LLL wave 
function for a textured state takes the form
\begin{equation}
\psi_{f, g}=\prod_i^N \left(\begin{array}{c} f(z_i)
\\ g(z_i) \\ \end{array}
\right)_i \psi_m \ .
\end{equation}
It is not hard to see by counting powers of $z$ that
the total physical
charge in these states is equal to the filling factor
times the maximum degree of 
$f$ and $g$ in $z$. 
A nicer argument, which makes the connection to the
$\sigma$ model manifest, is available for slowly varying textures. For
these one can use the plasma interpretation of the squared modulus of the
wavefunction to show that the local spin density and charge density of the 
quantum state are precisely the spin density and topological density of the
the $\sigma$ model solution and that its charge density is indeed just 
$\nu=1/m$ times the topological density of the texture.
Our assertion
about the charge of our wavefunctions then follows from the known
result for the net topological charge for the $\sigma$ 
model solutions.\cite{hole}

In particular, 
the variational wave function for a skyrmion with topological charge 
$n$, 
located at $z_0$ with core parameter $\lambda$ takes the form:\cite{moon} 
\begin{equation}
\psi_m^{z_0} =A\prod_i^N \left(\begin{array}{c} (z_i-z_0)^n 
\\ \lambda^n \\ \end{array}
\right)_i \psi_m,
\label{skyr}
\end{equation}
where, in $A$, we have included a normalization factor. $\lambda$ is a 
complex number; its
magnitude sets the core size and its phase determines
the internal spin orientation within the core region of the skyrmion.
This state carries extra positive charge (compared with the ground state)
near $z_0$ (i.e., the skyrmion
is hole-like since electrons carry negative charge), within spatial
range $|\lambda|$; spins are rotated away from the up direction in the
same region. In the limit $\lambda=0$ it reduces to the Laughlin 
quasihole wave function. The optimal core size is determined by the
competition between Zeeman energy and Coulomb exchange 
energy.\cite{sondhi,fertig} Wave functions for electron-like, anti-skyrmions can
be written down in the same spirit. In this paper we focus on hole like
skyrmions for convenience. The conclusions we reach, however, apply to
both skyrmions and anti-skyrmions.

The wavefunction for two skyrmions 
located at $z_a$ and $z_b$, with core parameters
$\lambda_a$ and $\lambda_b$, and winding numbers $n_a$ and $n_b$
is
\begin{equation}
\psi_m^{z_a, z_b} =A'\prod_i^N \left(\begin{array}{c} (z_i-z_a)^{n_a}
(z_i-z_b)^{n_b}
\\ \lambda_a^{n_a}(z_i-z_b)^{n_b}+\lambda_b^{n_b}(z_i-z_a)^{n_a} \\ \end{array}
\right)_i \psi_m.
\label{two}
\end{equation}
This is the appropriate two skyrmion wave function for as $z_i\sim z_a$
the wave function reduces to
\begin{equation}
\chi\sim \left(\begin{array}{c} (z_i-z_a)^{n_a}(z_i-z_b)^{n_b}
\\ \lambda_a^{n_a}(z_i-z_b)^{n_b} \end{array}\right)
\sim
\left(\begin{array}{c} (z_i-z_a)^{n_a} \\
\lambda_a^{n_a} \end{array}\right),
\end{equation}
which is the same as that
of a single skyrmion at $z_a$ with winding number $n_a$
and core parameter $\lambda_a$.
Similarly near $z_b$ the spinor wave function is identical to that of a
single skyrmion of winding number $n_b$ and core parameter $\lambda_b$.
An important difference between (\ref{two}) and (\ref{skyr}) is that
in (\ref{two}) the both up and
down spin components of the wave function depend on the location of the
skyrmions,
while in (\ref{skyr}) only up spin component depends on the location
of the skyrmion.\cite{note}

Finally, we record the explicit form of the wavefunction 
for a system with $M$ skyrmions, located at complex coordinates 
$\xi_1, \xi_2, \cdots, \xi_M$, with core parameters and winding numbers
$\lambda_1, \cdots, \lambda_M$ and $n_1, \cdots, n_M$ respectively:
\begin{equation}
\Psi=\prod_i^N\left[\prod_j^M(z_i-\xi_j)^{n_j}
\left(\begin{array}{c} 1 \\
\sum_k{\lambda_k^{n_k}\over (z_j-\xi_k)^{n_k}}\\ \end{array}\right)_i
\right]\psi_m.
\end{equation}
Clearly, as $z_i\sim \xi_j$, the wave function reduces to
\begin{equation}
\chi
\sim
\left(\begin{array}{c} (z_i-\xi_j)^{n_j} \\
\lambda_j^{n_j} \end{array}\right),
\end{equation}
which is that of a single skyrmion of the appropriate scale and winding number.

\noindent
{\bf Berry Phase Calculations:}
To determine the charge and statistics of the skyrmions, we take the
same approach used by Arovas {\em et al.} to determine the
same quantities for Laughlin quasiparticles and quasiholes.
We first consider the case where there is a single skyrmion in the system,
and consider the Berry phase for moving it around
a closed loop. This can be written as an integral over the skyrmion co-ordinate
$z_0$, 
\begin{equation}
\gamma=i\oint{dz_0}\langle\psi_m^{z_0}|
{\partial_{z_0}}\psi_m^{z_0}\rangle.
\end{equation}
 From Eq. (\ref{skyr}) we have
\begin{equation}
{\partial\psi_m^{z_0}\over \partial z_0}
=\left\{\sum_{j=1}^{N}\left(\begin{array}{cc}
{n\over z_0-z_j}
& 0 \\ 0 & 0\end{array} \right)_j\right\}\psi_m^{z_0},
\end{equation}
where the $2\times2$ matrix with subscript $j$ acts on the spinor wave 
function of the $j$th electron. We thus find that
\begin{equation}
\gamma=i\oint{dz_0}\left\langle\psi_m^{z_0}\left|\sum_j^N
\left(\begin{array}{cc}
{n\over z_0-z_j}
& 0 \\ 0 & 0\end{array} \right)_j\right|\psi_m^{z_0}\right\rangle.
\end{equation}
Recognizing that the average density of up spin electrons at point $z$ is
\begin{equation}
\rho_{\uparrow}^{z_0}(z)=\left\langle\psi_m^{z_0}\left|\sum_j
\left(\begin{array}{cc}
\delta(z_j-z)
& 0 \\ 0 & 0\end{array} \right)_j\right|\psi_m^{z_0}\right\rangle,
\end{equation}
we obtain
\begin{equation}
\gamma=i\oint{dz_0}\int{dxdy}\rho_{\uparrow}^{z_0}(z)
{n\over z_0-z}.
\label{phase}
\end{equation}
We write $\rho_{\uparrow}^{z_0}(z)
=\rho_{\uparrow}^0+\delta\rho_{\uparrow}^{z_0}(z)$,
where $\rho_{\uparrow}^0=1/(2\pi m\ell^2)$ 
is the density of up spin electrons in
the ground state, and $\delta\rho_{\uparrow}^{z_0}$ is the extra density of
up spin electrons due to the existence of the skyrmion itself.
If $z_0$ moves around a circle of radius $R$ clockwise,
the contribution to $\gamma$ due to the $\rho_{\uparrow}^0$ 
term is
\begin{equation}
\gamma_0=-2\pi n\langle N\rangle_R=-(2\pi n/m)(\Phi/\Phi_0),
\label{berry}
\end{equation}
where $\langle N\rangle_R$ is the average number of up spin electrons in
inside the circle, 
and $\Phi$ is the amount of magnetic flux enclosed by the
circle.  
The $\delta\rho$ term does not contribute to $\gamma$ 
due to the fact that $\delta\rho^{z_0}(z)$ is symmetric about the
point $z_0$.\cite{duncan} 
Since the phase picked up by a charged particle with charge $q$ 
after moving along a closed loop enclosing flux $\Phi$ is $-2\pi q\Phi/\Phi_0$,
we find from
Eq. (\ref{berry}) that the charge of a skyrmion described by the wave function
(\ref{skyr}) is $q=ne/m$, identical to that of the Laughlin quasiholes 
and independent of its core size, 
as expected.\cite{sondhi,moon,fertig}
This result has recently been obtained by Stone\cite{stone} using the effective
$\sigma$ model.

We now consider the situation where there are two
skyrmions located at $z_a$ and $z_b$, with core parameters
$\lambda_a$ and $\lambda_b$, winding numbers $n_a$ and $n_b$,
separated by a distance $R=|z_a-z_b|\gg|\lambda_{a,b}|$ and $R\gg\ell$, 
described by wave function (\ref{two}).
We let 
$z_a$ stand still,
and $z_b$ move along a closed loop enclosing 
$a$. We find that
the Berry phase picked up by the system is 
\begin{eqnarray}
\gamma&=&i\oint{dz_b}\langle\psi_m^{z_a,z_b}|
{\partial_{z_b}}\psi_m^{z_a,z_b}\rangle\nonumber\\
&=&im_b\oint{dz_b}\int{dxdy}\left[\rho_\uparrow(z){1\over z_b-z}
-\rho_\downarrow(z){\lambda_a^{m_a}(z-z_b)^{m_b-1}\over 
\lambda_a^{n_a}(z_i-z_b)^{n_b}+\lambda_b^{n_b}(z_i-z_a)^{n_a}}\right].
\end{eqnarray}
When the distance between the two skyrmions is much larger than the skyrmion
core sizes, we have $\rho_{\uparrow}(z)=\rho_{\uparrow}^0
+\delta\rho_{\uparrow}^{a}(z)
+\delta\rho_{\uparrow}^{b}(z)$, and $\rho_{\downarrow}(z)=
\delta\rho_{\downarrow}^a(z)+\delta\rho_{\downarrow}^b(z)$, 
where $\delta\rho_{\uparrow,(\downarrow)}^{a,(b)}$ are the 
extra charge density for up (down) spin electrons due to skyrmions $a$($b$),
which is non zero only near $z_a(z_b)$.
The $\rho_{\uparrow}^0$ term is identical to that in 
Eq. (\ref{berry}),
which is due to the charge of skyrmion $b$ moving in a magnetic field.
For the same reason as before, the $\delta\rho_{\uparrow}^{b}$ term vanishes.
Thus the additional phase due to the existence of skyrmion $a$ is 
\begin{equation}
\Delta\gamma=in_b\oint{dz_b}\int{dxdy}\left[
{\delta\rho_{\uparrow}^a(z)\over
z_b-z}-(\delta\rho_\downarrow^a(z)+\delta\rho_\downarrow^b(z))
{\lambda_a^{n_a}(z-z_b)^{n_b-1}\over
\lambda_a^{n_a}(z-z_b)^{n_b}+\lambda_b^{n_b}(z-z_a)^{n_a}}\right].
\end{equation}
In the large $R$ limit, $\delta\rho_{\uparrow,\downarrow}^{a,(b)}$ 
can be treated as $\delta$
functions localized at $z_a(z_b)$, with corrections to the above 
expression vanishing at least
as $1/R$. 
For $n_b>1$ or $n_b=1, n_a>1$
the $\delta\rho_{\downarrow}^b$ term vanishes in the large $R$
limit. Thus we obtain
\begin{equation}
\Delta\gamma=in_b\oint{dz_b}\int{dxdy}{\delta\rho_{\downarrow}^a(z)
+\delta\rho_{\uparrow}^a(z)\over z_b-z}=2\pi n_b q_a/e=2\pi n_an_b/m.
\end{equation}

This additional phase is due to the statistical interaction between the
skyrmions. We thus find that the statistical angle of the skyrmions,
which is the additional phase accumulated when two {\em identical}
hole like skyrmions with winding number $n$ are exchanged
clockwisely, to be 
\begin{equation}
\phi_h=\Delta\gamma/2=n^2\pi/m,
\label{stat}
\end{equation}
again identical to the Laughlin quasiholes and independent of the core
size (or spin).

Careful readers may worry at this point
that the variational wave functions (\ref{skyr}) and (\ref{two}) we use
here are not eigenstates of the total spin and its $z$ component which
are both good quantum numbers. On general grounds, we believe this is 
not a limitation on the validity of our results. The difference between
variational ``classical'' skyrmions and exact eigenstates is the
inclusion of symmetry related configurations. The Berry phases we compute
are {\em geometric} phases and will not be affected by the inclusion of
these additional configurations. For example, it is straightforward to 
project our variational wavefunctions onto states of definite $z$ component
of spin \cite{nayak,macd} and show that this does not change our results.

\noindent
{\bf Skyrmions with Topological Charge 1:}
There is a subtlety in our above considerations that we have glossed
over thus far. This is that in the special case of $n_a=n_b=1$, the 
large separation asymptotics used above in calculating the statistical 
phase $\delta\gamma$, no longer apply.
In contrast to the result for all $n > 1$, we find that $\Delta\gamma$
depends on the relative size and spin orientation of the two skyrmions
which is clearly unphysical for well separated skyrmions.

This problem can be traced to the power law tails of the variational
skyrmion wavefunctions, which are a consequence of the scale invariance
of the $O(3)$ $\sigma$-model. For $n > 1$, the power is large enough
that two skyrmions decouple at large distances but for $n=1$ they
never really do. 
Indeed, the single $n=1$ skyrmion wavefunction is pathological on its own;
the number of down spin particles in it (\ref{skyr})
{\em diverges} logarithmically with system size leading to a divergent
Zeeman energy in the presence of Zeeman field.\cite{nn} 

It is known that in the problem with Coulomb interactions, where their
competition with the Zeeman energy sets a scale for the skyrmions, the
form of the skyrmions is modified at large distances\cite{sondhi};
this issue will be further addressed elsewhere.\cite{ed}
What is important for our purposes is recognizing, that this will lead
to a decoupling of the $n=1$ skyrmions at large distances and
hence a result for their statistics consistent with those for the decoupled,
$n>1$ skyrmions (\ref{stat}) obtained above.

In summary, we have obtained the charge and statistical angle of quantum
Hall ferromagnet skyrmions by calculating Berry phases with
microscopic wave functions. We find they do not depend on the spin of the
skyrmions, and are identical to those of the Laughlin quasiholes at
the same filling factor. 

The authors would like to
thank R. N. Bhatt, S. M. Girvin, F. D. M. Haldane, 
A. H. MacDonald, K. Moon, V. Pasquier and S. C. Zhang 
for helpful discussions on this subject. This work was supported by NSF grant
DMR-9400362 (K.Y.).

\end{document}